# Topolectrical-circuit realization of 4D hexadecapole insulator

Weixuan Zhang[1*], Deyuan Zou[1*], Wenjing He[2], Jiacheng Bao[2], Qingsong Pei[1], Houjun Sun[2$], and Xiangdong Zhang[1+]

[1] Key Laboratory of advanced optoelectronic quantum architecture and measurements of Ministry of Education, Beijing Key Laboratory of Nanophotonics & Ultrafine Optoelectronic Systems, School of Physics, Beijing Institute of Technology, 100081, Beijing, China

[2] Beijing Key Laboratory of Millimeter wave and Terahertz Techniques, School of Information and Electronics, Beijing Institute of Technology, Beijing 100081, China

*These authors contributed equally to this work. +$Author to whom any correspondence should be addressed. E-mail: zhangxd@bit.edu.cn; sunhoujun@bit.edu.cn

**Recently, the theory of quantized dipole polarization has been extended to account for electric multipole moments, giving rise to the discovery of multipole topological insulators (TIs). Both two-dimensional (2D) quadrupole and three-dimensional (3D) octupole TIs with robust zero-dimensional (0D) corner states have been realized in various classical systems. However, due to the intrinsic 3D limitation, the higher dimensional multipole TIs, such as four-dimensional (4D) hexadecapole TIs, are supposed to be extremely hard to construct in real space, although some of their properties have been discussed through the synthetic dimensions. Here, we theoretically propose and experimentally demonstrate the realization of classical analog of 4D hexadecapole TI based on the electric circuits in fully real space. The explicit construction of 4D hexadecapole circuits, where the connection of nodes is allowed in any desired way free from constraints of locality and dimensionality, is provided. By direct circuit simulations and impedance measurements, the in-gap corner states protected by the quantized hexadecapole moment in the 4D circuit lattices are observed and the robustness of corner state is also demonstrated. Our work offers a new pathway to study the higher order/dimensional topological physics in real space.**

The exploration of topological physics in various systems[1-4] has become one of the most fascinating frontiers in recent years. Based on the bulk-boundary correspondence principle, the conventional topological phase is always featured by the boundary states with one-dimensional (1D) lower than the bulk that hosts them[5-11], e.g., 1D edge states of 2D quantum Hall systems and 2D spin surface states of 3D TIs. Recently, a novel class of symmetry-protected higher-order TIs that possess lower-dimensional boundary states have been proposed by generalizing the fundamental relationship between the Berry phase and quantized polarization, from dipole to multipole moments[12-27]. The most prominent examples are given by the $2^{th}$-order/2D quadrupole and $3^{th}$-order/3D octupole TIs, which possess robust 0D corner states protected by the quantized quadrupole/octupole polarizations. Motivated by the novel property, many experimental implementations of quadrupole and octupole TIs have been realized in various types of classical systems, including mechanics[13], acoustics[14-16], photonics[17, 18], and electrical circuits[19-21]. However, owing to the limitation of three spatial dimensions, the much higher order/dimensional multipole TIs, such as $4^{th}$-order/4D hexadecapole TIs, only seem of theoretical interest and still remain experimentally challenging to construct.

It is very attractive to explore topological physics beyond 3D, because there are many exotic topological phases in such higher dimensional systems. For instance, 4D/6D quantum Hall systems can exhibit a non-vanishing $2^{th}/3^{th}$ Chern number[28, 29] and 5D Weyl semimetals possess Yang monopoles and linked Weyl surfaces that are not shared by systems with three or fewer dimensions[30]. Recent advances in synthetic dimensions have provided an effective way to investigate higher dimensional physics in lower dimensionality[31-35], e.g., the dynamical version of the 4D quantum Hall effect has been fulfilled by two spatial dimensions plus two synthetic dimensions[34, 35]. However, the exploration of higher dimensional topology in fully real space has not been realized.

Recent investigations have shown that the property of electric circuit network depends only on how sites are connected, regardless of the shape of circuit lattice[36-39]. With such advantages of being versatile and reconfigurable, the higher dimensional electric circuits can always be projected into lower dimensional spaces with appropriate local/nonlocal site connections. Consequently, the circuit networks are regarded to be an ideal platform to construct higher dimensional topological phases in fully real space. On account of such superior performance, there are some theoretical schemes for creating higher dimensional topological matters using electric circuits, for example, the $n^{th}$-Chern insulator in $2n$-dimensional space[40], 4D TIs in class AI[41] and Seifert surface within 3D boundary states of 4D nodal systems[42] have been fulfilled using circuit networks. While, the experimental demonstration of probing higher-dimensional topology using electric circuit is still lacking.

Here, we theoretically propose and experimentally demonstrate the realization of $4^{th}$-order/4D hexadecapole TI using electric circuits. The explicit construction of 4D hexadecapole circuit, which are not limited by the 3D constraints and possess tunable site connections, is provided. The in-gap corner states protected by the quantized hexadecapole moment are directly observed in the 4D circuit lattices through impedance measurements. Additionally, the robustness of corner state is also demonstrated with two types of on-site perturbations. We except that the method of creating multipole TIs using electric circuits can be extended to the much higher dimensions and provide a reference for the study of higher-order/dimensional topological physics with the help of flexible electrical circuits.

**Hexadecapole topological insulator and its circuit realization.** The most important feature of multipole TIs is their cascade hierarchy of topology. This property indicates that the $(n-1)^{th}$-order multipole TIs are the projection of the $n^{th}$-order multipole TIs into the space with 1D reduction. Conversely, the $n^{th}$-order multipole TIs are also able to be constructed by suitably connecting a pair of $(n-1)^{th}$-order multipole TIs with opposite signs[12]. Hence, it is straightforward to infer that the unit cell of 4D hexadecapole TIs can be directly constructed by linking a pair of 3D octupole TIs with opposite settings of the site connection. The schematic diagram is plotted in the left chart of Fig. 1a, where the solid and dash (blue/green/purple/yellow) lines represent the positive and negative intra-cell couplings (along $x$-axis/$y$-axis/$z$-axis/$w$-axis), respectively. For intuitive illustration, we provide an alternative graphic representation of the proposed 4D tight-binding lattice model in the 2D plane, as shown in the right chart of Fig. 1a. The unit cell contains sixteen sites marked by 1…16 and each site is connected with other four sites in different directions, forming the intra-cell coupling ($\gamma$) pattern. By reasonably arranging the units along eight directions ($\pm x$, $\pm y$, $\pm z$, $\pm w$) in the four spatial dimensions, the whole lattice model of the hexadecapole TI can be realized. To manifest the manner of inter-cell couplings ($\lambda$), the corresponding 3D projection of the 4D unit cell should be clarified (See Supporting Information A for details). In this case, by suitably tuning the ratio between intra-cell and inter-cell couplings ($\gamma/\lambda$), the non-trivial hexadecapole TIs can be realized.

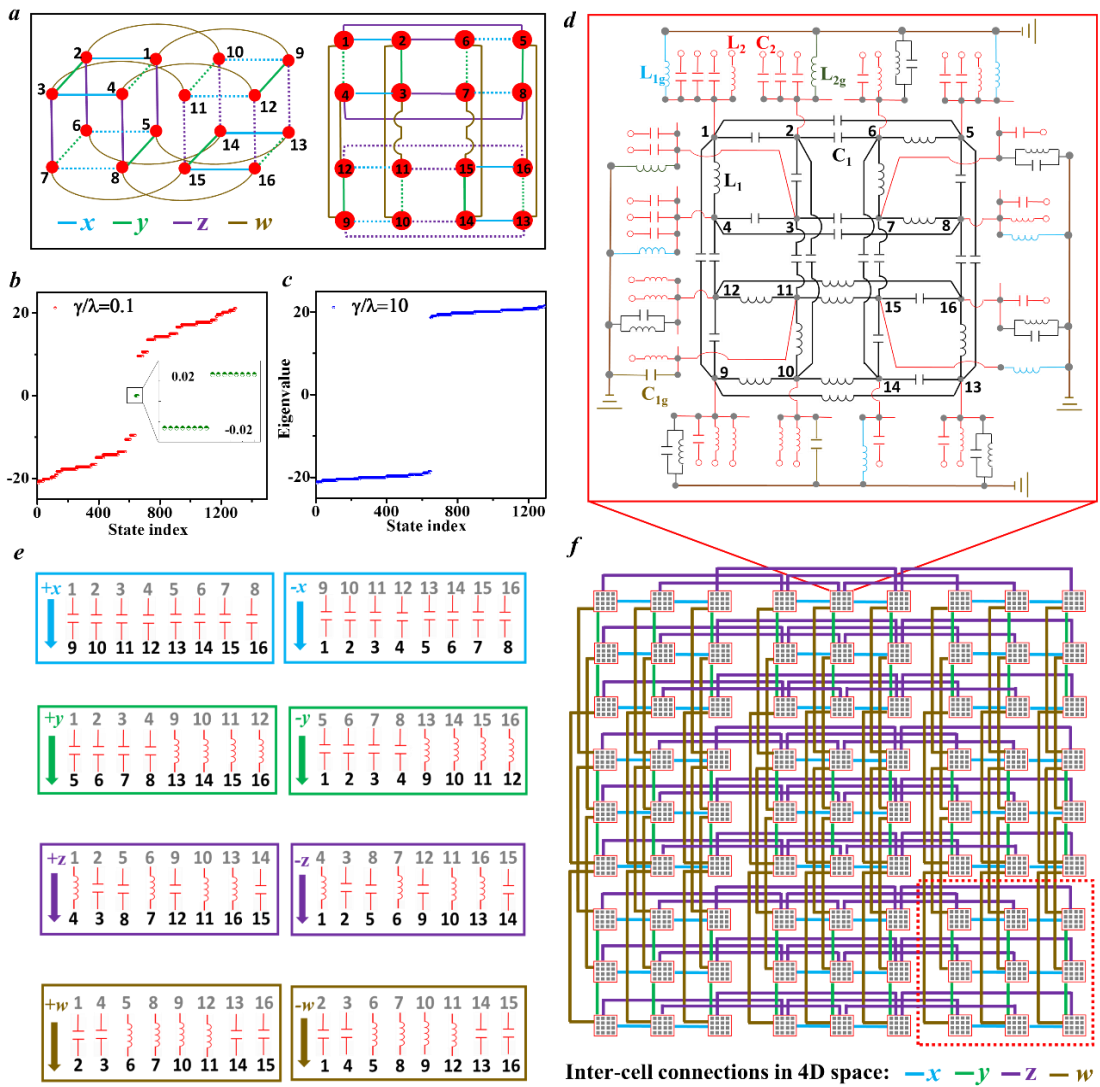

**Figure 1 | Hexadecapole topological insulator and its circuit realization. a**, Left panel illustrates the schematic of constructing the lattice model with quantized hexadecapole moments by linking a pair of

octupole insulators with opposite signs. The right panel plots the 2D projection of tight-binding model of 4D hexadecapole topological insulator. **b** and **c**, The energy spectra of a finite $3\times3\times3\times16$ lattice model with the ratio between the intra- and inter-cell coupling being $\gamma/\lambda=0.1$ and $\gamma/\lambda=10$, respectively. The inset plots the enlarged view of corner states in the non-trivial phase. The energy splitting is due to the finite size effect. **d**, The electric circuit realization of the model in **a**. Here the unit cell contains sixteen nodes coupled by two pairs of capacitors and inductors (same resonance frequency) corresponding to the intra-cell, ($L_1$, $C_1$) in black, and inter-cell, ($L_2$, $C_2$) in red, couplings. The grounding of each site is suitably designed for realizing identical onsite potential. **e**, The manner of inter-cell site-connections along eight directions in 4D space. **f,** Layout of the projected 2D circuit from 4D lattice ($3\times3\times3\times16$). The blue/green/purple/yellow connections represent the site coupling along the x-axis/y-axis/z-axis/*w*-axis.

The above proposed 4D lattice model can be effectively described by a tight-binding Hamiltonian. In Bloch representation, the Hamiltonian can be expressed as:

$$H(k_{4D}) = (\gamma+\lambda\cos k_x)\Gamma_1 + \lambda\sin k_x\Gamma_2 + (\gamma+\lambda\cos k_y)\Gamma_3 + \lambda\sin k_y\Gamma_4 \\ + (\gamma+\lambda\cos k_z)\Gamma_5 + \lambda\sin k_z\Gamma_6 + (\gamma+\lambda\cos k_w)\Gamma_7 + \lambda\sin k_w\Gamma_8 \quad (1)$$

where $\Gamma_1=\sigma_1\otimes I_{8\times8}$, $\Gamma_2=-\sigma_2\otimes I_{8\times8}$, $\Gamma_3=\sigma_3\otimes\sigma_1\otimes I_{4\times4}$, $\Gamma_4=-\sigma_3\otimes\sigma_2\otimes I_{4\times4}$, $\Gamma_5=\sigma_3\otimes\sigma_3\otimes\sigma_2\otimes\sigma_2$, $\Gamma_6=\sigma_3\otimes\sigma_3\otimes\sigma_1\otimes\sigma_2$, $\Gamma_7=\sigma_3\otimes\sigma_3\otimes I_{2\times2}\otimes\sigma_1$ and $\Gamma_8=-\sigma_3\otimes\sigma_3\otimes\sigma_3\otimes\sigma_2$. $\sigma_i$ (i=1, 2, 3) are three Pauli matrices and $I_{j\times j}$ is the *j*-dimensional unitary matrix (See Supporting Information B for details). This 16-band Bloch Hamiltonian possesses four reflection symmetries $M_i H(k_{4D}) M_i^\dagger = H(m_i k_{4D})$ for *i=x, y,* z*, w*, where $M_x=\sigma_1\otimes\sigma_3\otimes I_{2\times2}\otimes\sigma_3$, $M_y=I_{2\times2}\otimes\sigma_1\otimes I_{2\times2}\otimes\sigma_3$, $M_z=I_{4\times4}\otimes\sigma_1\otimes\sigma_1$, $M_w=I_{4\times4}\otimes\sigma_3\otimes\sigma_1$ and $m_x(k_x,k_y,k_z,k_w)=(-k_x,k_y,k_z,k_w)$, $m_y(k_x,k_y,k_z,k_w)=(k_x,-k_y,k_z,k_w)$, $m_z(k_x,k_y,k_z,k_w)=(k_x,k_y,-k_z,k_w)$, $m_w(k_x,k_y,k_z,k_w)=(k_x,k_y,k_z,-k_w)$. These reflection operators obey $\{M_i, M_j\}=0$, for $i, j = x, y,$ z, *w* and $i\neq j$. The presence of four anti-commute reflection symmetries indicates that the hexadecapole moment can only take quantized values of 0 or $\pm e/2$ at half filling, depending on the ratio between the intra- and inter-cell couplings ($\gamma/\lambda$). Similar to the quadrupole and octupole TIs, the topology of our proposed 4D hexadecapole TIs can also be evaluated using the nested Wilson loop, and the value of Wannier sector polarization $P_\alpha$ can be given along four spatial dimensions ($\alpha=x, y,$ z, *w*) (See Supporting Information C for details). When the ratio $\gamma/\lambda$ is taken from -1 to 1, the periodic lattice model possesses the nontrivial hexadecapole moment

$(p_x,p_y,p_z,p_w)=(1/2,1/2,1/2,1/2)$, which leads to the existence of mid-gap 0D corner states in the corresponding open lattice. In contrast, the model goes into trivial phase, $(p_x,p_y,p_z,p_w)=(0,0,0,0)$, with $|\gamma/\lambda|>1$. Fig. 1b shows the calculated energy spectrum of a finite $3\times3\times3\times16$ lattice model (totally 1296 sites) with smaller intra-cell couplings ($\gamma/\lambda=0.1$). It is found that sixteen $4^{th}$-order 0D corner states, marked by the green dots, appear within the energy gap. And, the energy splitting of corner states (presented in the inset of Fig. 1b) is due to the finite size effect of the system, where the hybridization of corner states at different locations exists. For comparison, we also calculate the energy spectrum with larger intra-cell couplings ($\gamma/\lambda=10$), as shown in Fig. 1c. In this case, the in-gap states disappear owing to the trivial topology.

In the following, we discuss the experimentally feasible scheme for the above proposed hexadecapole TIs using electric circuits. Similar to the representation of 4D lattice model in the 2D plane, the 4D unit cell of the electric circuit can also be designed in the 2D plane with versatile circuit connections, as shown in Fig. 1d. The unit cell contains sixteen nodes labeled from 1 to 16, corresponding to the 16 sites in the tight-binding lattice mode. Two pairs of capacitors and inductors [$(L_1, C_1)$ in black and $(L_2, C_2)$ in red], which possess the same resonance angular frequency $\omega_0=1/(L_1C_1)^{1/2}=1/(L_2C_2)^{1/2}$ and satisfy the relationships of $C_2=\kappa C_1$ and $L_1=\kappa L_2$, are used to simulate the intra- and inter-cell couplings. The positive and negative couplings are achieved by the interconnected capacitors and inductors, respectively. By suitably arranging the unit cell along four spatial dimensions, the periodic 4D electric circuits can be realized. The method for constructing the inter-cell coupling in electric circuits is identical with the corresponding 4D tight-binding lattice model. Fig. 1e shows the inter-cell coupling pattern, where eight insets illustrate the nodes-connection of the adjacent units along $\pm x$-axis, $\pm y$-axis, $\pm z$-axis and $\pm w$-axis (arrows mark the linking directions), respectively. As for the grounding setting, to ensure these sites have identical resonance frequency (equivalent to the same on-site potential of the lattice model), sites 1, 4, 5, 8, 14 and 15 are grounded with inductor $L_{1g}=L_1/(2+2\kappa)$, sites 2 and 3 are grounded with inductor $L_{2g}=L_1/(4+4\kappa)$, sites 10 and 11 are connected to the ground via the capacitor $C_{1g}=2(C_1+C_2)$ and sites 6, 7, 9, 12, 13 and 16 are connected to the ground via an LC circuit with capacitor and inductor being $C_1$ and $L_1$. Through the appropriate setting of circuit's grounding and connecting, the designed periodic 4D circuit can exhibit nontrivial hexadecapole moments at the resonance frequency $\omega_0$.

To clarify the symmetry of the proposed translation invariant circuit, we derive the 16×16 circuit Laplacian $\tilde{J}(\omega_0, k_{4D})$ in the moment space at $\omega_0$ based on the Kirchhoff's current law (See Supporting Information D for details). It can be written as:

$$\tilde{J}(\omega_0, k_{4D}) = -i\sqrt{C_1/L_1}\,[(1+\kappa\cos k_x)\Gamma_1 + \kappa\sin k_x \Gamma_2 + (1+\kappa\cos k_y)\Gamma_3 + \kappa\sin k_y \Gamma_4 \\ + (1+\kappa\cos k_z)\Gamma_5 + \kappa\sin k_z \Gamma_6 + (1+\kappa\cos k_w)\Gamma_7 + \kappa\sin k_w \Gamma_8]. \quad (2)$$

Here, $\Gamma_i$ (i=1, 2, 3, 4, 5, 6, 7, 8) possess the same form used in Eq. (1). In this case, the circuit Laplacian, taking a role similar to the Hamiltonian, also satisfies four anti-commute reflection symmetries, which are written as:

$$\begin{aligned} M_x J_\lambda(\omega_0, q_x, q_y, q_z) M_x^\dagger &= J_\lambda(\omega_0, -k_x, k_y, k_z, k_w) \\ M_y J_\lambda(\omega_0, q_x, q_y, q_z) M_y^\dagger &= J_\lambda(\omega_0, k_x, -k_y, k_z, k_w) \\ M_z J_\lambda(\omega_0, q_x, q_y, q_z) M_z^\dagger &= J_\lambda(\omega_0, k_x, k_y, -k_z, k_w) \\ M_w J_\lambda(\omega_0, q_x, q_y, q_z) M_w^\dagger &= J_\lambda(\omega_0, k_x, k_y, k_z, -k_w) \end{aligned} \quad (3)$$

It is found that our proposed electric circuit possesses identical reflection symmetries as the Bloch Hamiltonian of the above tight-binding lattice model. Hence, if $\kappa \neq 1$ the spectrum of $\tilde{J}(\omega_0, k_{4D})$ is gapped. In this condition, when the intra-cell coupling is smaller than the inter-cell coupling ($\kappa > 1$) the electric circuit can be regarded as the classical analog of the hexadecapole TIs. On the contrary, if the intra-cell coupling is larger than the inter-cell coupling ($\kappa < 1$), the circuit becomes a trivial case.

Now, we turn to the 4D electric circuit with open boundaries to realize the 0D corner states protected by the non-zero hexadecapole moment. Here, the open circuit with 3×3×3×3 units is considered. As presented in Fig. 1f, the finite 4D circuit is projected into a 2D plane with non-local site connections, where the manner of inter-cell connections is the same as the periodic circuit (displayed in Fig. 1e). The little red square (containing 16 nodes) represents the unit of circuit with appropriate intra-cell couplings (as plotted in Fig. 1d). The 3×3 array of this unit cell (enclosed by the red dash wireframe) illustrates the inter-cell coupling within the *x-y* plane, where the green (blue) lines present the coupling along the *x*-axis (*y*-axis). Additionally, the large 3×3 array of red dash wireframes manifests the z-*w* space. The linked purple (yellow) lines correspond to the inter-cell coupling along the z-axis (*w*-axis). In this case, the 4D circuit model can be constructed in the 2D plane with appropriate non-local node-connections. Furthermore, it is worth noting that the grounding of finite circuits should also be suitably designed to ensure the same resonance frequency at each circuit nodes (the identical on-site energy) (see Supporting Information E for details).

**Theoretical results for the Hexadecapole topological circuit.** To prove the validity of the designed hexadecapole topological circuit ($3\times3\times3\times3$), we calculate the spectrum of circuit Laplacian in real space as a function of the driving frequency ($f=\omega/2\pi$), as shown in Fig. 2a. In the calculation, the values of $C_1$ ($L_1$) and $C_2$ ($L_2$) are taken as 1nF (3.3uH) and 10nF (0.33uH), respectively. In this case, the coupling ratio between the intra-cell and inter-cell equals to $\kappa=0.1$ ($\gamma/\lambda=0.1$). It is clearly seen that the isolated state appears in the spectrum gap (red region) around the resonance frequency $f_0=2.77$MHz. This state corresponds to the 0D corner mode induced by the non-trivial hexadecapole moment. When the driving frequency is away from the resonance condition ($f\neq f_0$), the circuit Laplacian (with heterogeneous onsite potentials) no longer reproduces the tight-binging lattice model with non-trivial hexadecapole moments, and the in-gap corner states also gradually disappear. The distribution of 'zero-energy' modes of $J(\omega_0, r_{4D})$ is plotted in Fig. 2b with single-site resolution. It is clearly shown that the modal profile is spatially concentrated at 16 nodes (corner positions of the 4D lattice) in the open circuit, which manifests the existence of 0D corner state.

For further validation of our design, we also perform steady-state simulations of the proposed open circuit using the LTspice software. The calculated frequency-dependent impedance (normalized with respect to the maximum) between two nearest-neighbor sites on different positions are shown in Fig. 2c. The red/blue, green, pink and black lines correspond to the results with node locating in the 4D/3D bulk, 2D surface, 1D edge and 0D corner of the finite circuit, respectively. It is noted that the bulk, surface, and edge sites sustain lower impedances in a wide frequency range. Meanwhile, the impedance is extremely high at the corner (around $f=f_0$, away from other states), indicating the existence of in-gap 0D corner state. The frequency splitting is result from the finite size effect. For comparison, we also calculate the frequency-dependent impedance response on different positions of the trivial circuit (larger intra-cell coupling), where $C_1$ ($L_1$) and $C_2$ ($L_2$) are taken as 10nF (0.33uH) and 1nF (3.3uH), that is $\kappa=10$ ($\gamma/\lambda=10$), as presented in Fig. 2d. In this case, no corner state can be observed in the trivial circuit. The above discussions only focus on the $3\times3\times3\times3$ array. In fact, we have also calculated impedance responses for the sample of other sizes, where the similar phenomena can always be observed.

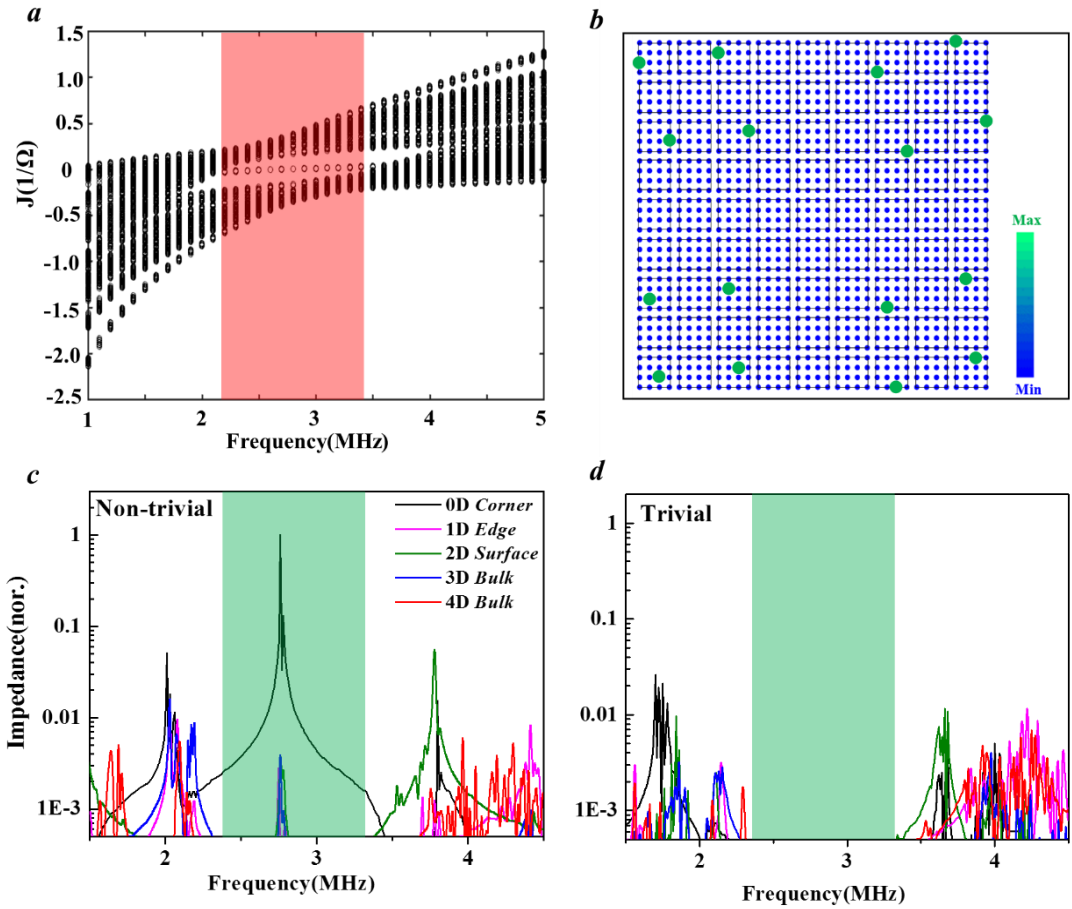

**Figure 2 | Theoretical results for the circuit spectrum and corner mode. a,** Theoretical spectrum of the circuit Laplacian in real space $J(\omega,r_{4D})$ as a function of the driving frequency. The isolated mode crossing the gap, which corresponds to a zero-energy eigenvalue of circuit Laplacian $J(\omega_0,r_{4D})$, is clearly visible. It corresponds to the topological corner mode. **b,** The distribution of impedance profile for the zero-energy mode at $f=f_0$. Sixteen sites with large impedances manifest the existence of 0D corner state. **c and d,** Steady-state simulations of the $(3\times3\times3\times3)$ open circuit with non-trivial and trivial hexadecapole

moments. The red/blue, green, pink and black lines correspond to results with the site locating in 4D/3D bulk, 2D surface, 1D hinge and 0D corner, respectively. Here, $C_1$ ($L_1$) and $C_2$ ($L_2$) are taken as 10nF (0.33uH) and 1nF (3.3uH) and the serial resistance of inductors is set as 1mΩ.

**Experimental demonstration of the circuit hexadecapole topological insulator.** To experimentally observe the 0D corner state induced by the 4D hexadecapole moment, the designed electric circuit with 3×3×3×3 units (the same to the above theoretical model) is fabricated. The photograph image of the unit cell is shown in the top chart of Fig. 3a. Similar to the theoretical model, there are sixteen nodes in the unit cell connected through $L_1$ (3.3uH) and $C_1$ (1nF), forming the intra-cell coupling pattern. And, the inter-cell site-connections are realized with $L_2$ (0.33uH) and $C_2$ (10nF). It is worthy to note that the tolerance of the circuit elements is only 1% to avoid the detuning of corner resonance. In this case, the resonant frequency of the circuit is 2.77MHz. For the construction of 4D hexadecapole electric circuit in the 2D space, totally nine printed circuit boards (PCBs) with non-local site-connections are applied, as shown in the bottom chart of Fig. 3a. Each sub-PCB contains 3×3 units, which are suitably connected to form the inter-cell couplings along *x*-axis and *y*-axis. The non-local site connections between the units of different sub-PCBs perform the inter-cell couplings along z-axis (red lines) and *w*-axis (black lines). With suitably local and non-local inter-cell couplings, the hexadecapole topological circuit is constructed in the 2D space.

We use Wayne kerr precision impedance analyzer to measure the impedance of the circuit as a function of the driving frequency. The experimental results are shown in Fig. 3b. The black line represents the measured impedance of the site located at the corner position. We find that the extremely high impedance (450Ω for the peak value) appears at the resonance frequency, resulting from the existence of the corner mode. Comparing with the numerical simulation (Figure 2c), the wider peak of the measured result is mainly due to the larger serial resistance of inductors (about 1000mΩ) and resistive loss of linking wires. In addition, the red, blue, green and pink lines present the measured impedance of the site locating at 4D bulk, 3D bulk, 2D surface and 1D hinge, respectively. It is shown that the impedances on these positions are relative small in a broad spectra range. For comparison purposes, we also fabricated a circuit with trivial bulk property (intra-cell coupling larger than the inter-cell coupling, $\kappa<1$) but same resonant frequency (2.77MHz), where $C_1$ ($L_1$) and $C_2$ ($L_2$) are chosen as 10nF (0.33uH) and 1nF (3.3uH). In such a case, the measured impendences on different positions are plotted in Fig. 3c. We note that no corner state can be observed in such a trivial circuit. This result further demonstrates that the existence of corner mode is not owing to the local effects particular to the physical boundaries of the array and it is result from the non-trivial hexadecapole moment of the bulk.

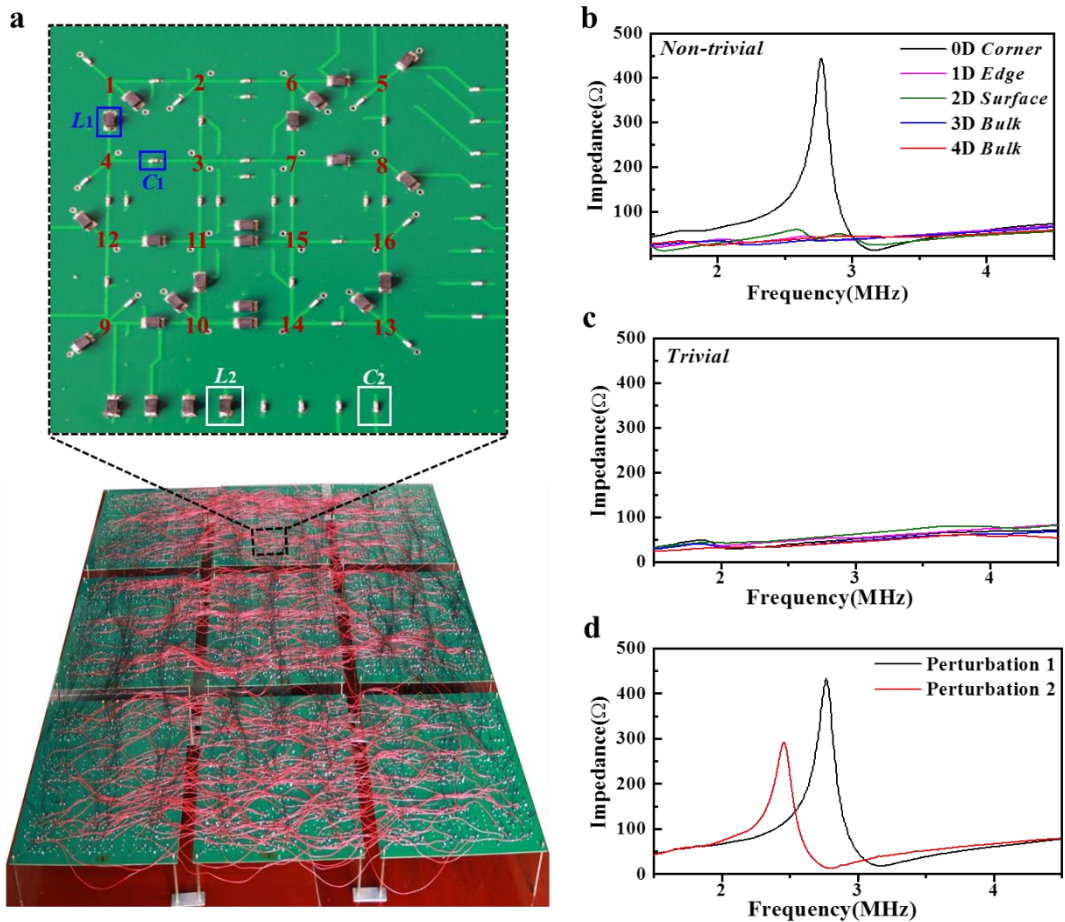

**Figure. 3 | Experimental demonstration of the circuit hexadecapole topological insulator. a,** The photo of the unit cell and whole topological circuit, projected from the original 4D circuit, with non-trivial hexadecapole moments. There are sixteen nodes in the unit cell connected through $L_1$ (3.3uH) and $C_1$ (1nF), forming the intra-cell coupling pattern. And, the inter-cell site-connections are realized with $L_2$ (0.33uH) and $C_2$ (10nF). The whole sample contains totally nine printed circuit boards (PCBs). The inter-cell connections within the same (different) sub-PCB form the inter-cell couplings along *x*- and *y*-axis (*z*- and *w*-axis). **b and c,** Measured frequency-dependent impedance of the nontrivial and trivial circuits on different locations. The red/blue, green, pink and black lines correspond to the result with site locating in 4D/3D bulk, 2D surface, 1D hinge and 0D corner, respectively. **d,** Experimentally measured spectra at

the corner site with two types of perturbations. The black (red) line corresponds to the resonance frequency shifts (adding a grounding capacitor 1nF) on the four sites linked to the corner (corner site).

**Discussion.**

The most important property of the topologically protected corner state is that it is robust against defects. To demonstrate this effect, two types of perturbation are introduced: resonance frequency shifts on the four sites linked to the corner location (perturbation 1) and resonance frequency shift on the corner site (perturbation 2). Here, we utilize a simple method to induce the resonant frequency shift on the perturbed sites by adding an extra capacitor on its grounding part. When the values of the perturbed capacitors change from 0nF to 10nF, our simulation results for the circuit Laplacian and impedance response show that the corner state is almost unaffected for the case of perturbation 1, the resonance peak still exists but is shifted to the other frequency for the case of perturbation 2. The detailed results are given in Supporting Information F. The corresponding experimental results are also consistent with the circuit simulations. The black and red lines in Fig. 3d represent the experimentally measured results under the two types of perturbation, respectively. In such a case, the 1nF capacitors are added on the perturbed sites of the fabricated sample. It is clearly seen that the resonance peaks still exist under the perturbations, which are identical with the circuit simulations, manifesting the robustness of the corner mode.

In conclusion, we have presented an observation of a new class of TIs with quantized hexadecapole moments in fully real space. The in-gap corner states protected by the quantized hexadecapole moment have been observed by direct circuit simulations and impedance measurements, and the robust properties of these corner states have also been demonstrated. In practice, the produced sixteen corner states in 2D space projected from 4D hexadecapole topological circuits can be used to realize robust trapping of electromagnetic energy at multiple locations. Moreover, by replacing capacitors in the LC network with varicap diodes, whose linearized capacitance depends on the applied bias voltage, the topological transition in a hexadecapole TIs can also be observed. The design principle presented in this work is universal. According to such a method, any higher dimensional multiple topolectrical-circuit insulator can be constructed with more complicated nonlocal circuit connections. The method can also be extended to non-Hermitian higher dimensions topological circuits by using negative impedance. We expect that the experimental access to higher order/dimensional topological physics with electric circuit in fully real space can provide a valuable reference for many other directions of research.

## Methods

**Sample fabrications.** We exploit the electric circuits by using PADs program software, where the PCB composition, stackup layout, internal layer and grounding design are suitably engineered. Here, each well-designed PCB possesses totally ten layers to arrange the site-couplings (intra-cell coupling and inter-cell couplings along x-axis and y-axis). It is worthy to note that the ground layer should be placed in the gap between any two layers to avoid their coupling. Moreover, all PCB traces have a relatively large width (0.5mm) to reduce the parasitic inductance and the spacing between electronic devices is also large enough (1.0mm) to avert spurious inductive coupling. On the other hand, the inter-cell couplings along z-axis and $w$-axis are achieved through the non-local site connections between totally nine sub-PCBs. For the convenience of experimental measurements, ten SMP connectors are welded on 0D corner, 1D edge, 2D surface, 3D/4D bulk of the whole PCB. Additionally, to ensure the same grounding condition, we link the copper pillar of each sub-PCB together.

## Acknowledgements

This work was supported by the National key R & D Program of China under Grant No. 2017YFA0303800 and the National Natural Science Foundation of China (No.91850205 and No.61421001).

## Author contributions

W. X. Zhang and D. Y. Zou finished the theoretical scheme and designed the experiments. D. Y. Zou, W. X. Zhang, W. J. He, J. C. Bao and Q. S. Pei finished experiments under the supervision of H. J. Sun and X. D. Zhang. X. D. Zhang initiated and designed this research project.

## Competing interests

The authors declare no competing interests.

# Supporting Information

# Topolectrical-circuit realization of 4D hexadecapole insulator

Weixuan Zhang[1*], Deyuan Zou[1*], Wenjing He[2], Jiacheng Bao[2], Qingsong Pei[1], Houjun Sun[2$],
and Xiangdong Zhang[1+]

[1]Key Laboratory of advanced optoelectronic quantum architecture and measurements of Ministry of Education, Beijing Key Laboratory of Nanophotonics & Ultrafine Optoelectronic Systems, School of Physics, Beijing Institute of Technology, 100081, Beijing, China

[2] Beijing Key Laboratory of Millimeter wave and Terahertz Techniques, School of Information and Electronics, Beijing Institute of Technology, Beijing 100081, China

*These authors contributed equally to this work. +$Author to whom any correspondence should be addressed. E-mail: zhangxd@bit.edu.cn; sunhoujun@bit.edu.cn

## A. 3D projection of 4D hexadecapole Topological insulators

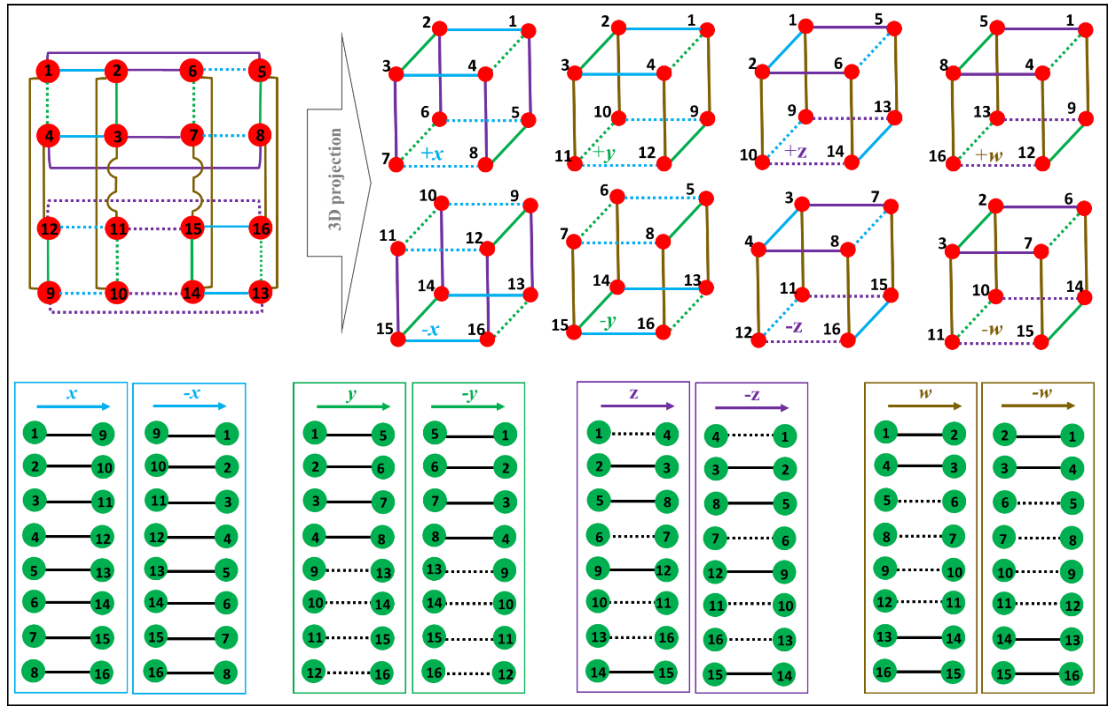

**Figure S1 | The 3D projection of 4D hexadecapole topological insulators.** The unit cell of the hexadecapole topological insulator with ideal intra-cell coupling is presented in the up-left inset. The blue/green/purple/yellow lines represent the site-coupling parallel to *x/y/z/w* directions. The projected 3D octupole TIs are presented by eight cubes. The bottom inset plots the pattern of inter-cell coupling along eight directions.

To manifest the inter-cell couplings of the hexadecapole TIs, the corresponding 3D projection of 4D hexadecapole TIs into different bulks (boundaries of 4D space) should be clarified. As shown in Fig. S1, we plot four pairs of 3D octupole TIs, which are derived from four orthogonal projection directions, respectively. Based on these projected 3D octupole TIs, the form of inter-cell couplings between adjacent units along different directions (+x, -x, +y, -y, +z, -z, +w, -w) is obtained, as plotted in the bottom inset of Fig. S1. By suitably arranging the units along eight directions ($\pm x, \pm y, \pm z, \pm w$) in the four spatial dimensions, the whole lattice model of the hexadecapole TI can be realized.

## B. Tight-binding lattice model of the hexadecapole Topological insulators

The tight-binding Hamiltonian of our proposed hexadecapole topological insulators can be expressed as:

$$H(x, y, z, w) = H_{intra}(x, y, z, w) + H_{inter}(x, y, z, w),  \tag{S1}$$

where the intra- and inter-cell coupling terms can be written as:

$$\begin{aligned}
H_{intra} = \sum_{x,y,z,w} \gamma &\{[c^\dagger_{x,y,z,w}(1)c_{x,y,z,w}(2) - c^\dagger_{x,y,z,w}(1)c_{x,y,z,w}(4) + c^\dagger_{x,y,z,w}(1)c_{x,y,z,w}(5) + c^\dagger_{x,y,z,w}(1)c_{x,y,z,w}(9)] \\
&+ [c^\dagger_{x,y,z,w}(2)c_{x,y,z,w}(3) + c^\dagger_{x,y,z,w}(2)c_{x,y,z,w}(6) + c^\dagger_{x,y,z,w}(2)c_{x,y,z,w}(10)] \\
&+ [c^\dagger_{x,y,z,w}(3)c_{x,y,z,w}(4) + c^\dagger_{x,y,z,w}(3)c_{x,y,z,w}(7) + c^\dagger_{x,y,z,w}(3)c_{x,y,z,w}(11)] \\
&+ [c^\dagger_{x,y,z,w}(4)c_{x,y,z,w}(8) + c^\dagger_{x,y,z,w}(4)c_{x,y,z,w}(12)] \\
&+ [-c^\dagger_{x,y,z,w}(5)c_{x,y,z,w}(6) + c^\dagger_{x,y,z,w}(5)c_{x,y,z,w}(8) + c^\dagger_{x,y,z,w}(5)c_{x,y,z,w}(13)] \\
&+ [-c^\dagger_{x,y,z,w}(6)c_{x,y,z,w}(7) + c^\dagger_{x,y,z,w}(6)c_{x,y,z,w}(14)] \\
&+ [-c^\dagger_{x,y,z,w}(7)c_{x,y,z,w}(8) + c^\dagger_{x,y,z,w}(7)c_{x,y,z,w}(15)] + c^\dagger_{x,y,z,w}(8)c_{x,y,z,w}(16) \\
&+ [-c^\dagger_{x,y,z,w}(9)c_{x,y,z,w}(10) + c^\dagger_{x,y,z,w}(9)c_{x,y,z,w}(12) - c^\dagger_{x,y,z,w}(9)c_{x,y,z,w}(13)] \\
&+ [-c^\dagger_{x,y,z,w}(10)c_{x,y,z,w}(11) - c^\dagger_{x,y,z,w}(10)c_{x,y,z,w}(14)] \\
&+ [-c^\dagger_{x,y,z,w}(11)c_{x,y,z,w}(12) - c^\dagger_{x,y,z,w}(11)c_{x,y,z,w}(15)] - c^\dagger_{x,y,z,w}(12)c_{x,y,z,w}(16) \\
&+ [c^\dagger_{x,y,z,w}(13)c_{x,y,z,w}(14) - c^\dagger_{x,y,z,w}(13)c_{x,y,z,w}(16)] \\
&+ c^\dagger_{x,y,z,w}(14)c_{x,y,z,w}(15) + c^\dagger_{x,y,z,w}(15)c_{x,y,z,w}(16)\} + H.C.
\end{aligned}  \tag{S2}$$

$$\begin{aligned}
H_{inter} = \sum_{x,y,z,w} \lambda &\{[c^\dagger_{x,y,z,w}(1)c_{x+1,y,z,w}(9) + c^\dagger_{x,y,z,w}(2)c_{x+1,y,z,w}(10) + c^\dagger_{x,y,z,w}(3)c_{x+1,y,z,w}(11) + c^\dagger_{x,y,z,w}(4)c_{x+1,y,z,w}(12) \\
&+ c^\dagger_{x,y,z,w}(5)c_{x+1,y,z,w}(13) + c^\dagger_{x,y,z,w}(6)c_{x+1,y,z,w}(14) + c^\dagger_{x,y,z,w}(7)c_{x+1,y,z,w}(15) + c^\dagger_{x,y,z,w}(8)c_{x+1,y,z,w}(16)] \\
&+ [c^\dagger_{x,y,z,w}(1)c_{x,y+1,z,w}(5) + c^\dagger_{x,y,z,w}(2)c_{x,y+1,z,w}(6) + c^\dagger_{x,y,z,w}(3)c_{x,y+1,z,w}(7) + c^\dagger_{x,y,z,w}(4)c_{x,y+1,z,w}(8) \\
&- c^\dagger_{x,y,z,w}(9)c_{x,y+1,z,w}(13) - c^\dagger_{x,y,z,w}(10)c_{x,y+1,z,w}(14) - c^\dagger_{x,y,z,w}(11)c_{x,y+1,z,w}(15) - c^\dagger_{x,y,z,w}(12)c_{x,y+1,z,w}(16)] \\
&+ [-c^\dagger_{x,y,z,w}(1)c_{x,y,z+1,w}(4) + c^\dagger_{x,y,z,w}(2)c_{x,y,z+1,w}(3) + c^\dagger_{x,y,z,w}(5)c_{x,y,z+1,w}(8) - c^\dagger_{x,y,z,w}(6)c_{x,y,z+1,w}(7) \\
&+ c^\dagger_{x,y,z,w}(9)c_{x,y,z+1,w}(12) - c^\dagger_{x,y,z,w}(10)c_{x,y,z+1,w}(11) - c^\dagger_{x,y,z,w}(13)c_{x,y,z+1,w}(16) + c^\dagger_{x,y,z,w}(14)c_{x,y,z+1,w}(15)] \\
&+ [c^\dagger_{x,y,z,w}(1)c_{x,y,z,w+1}(2) + c^\dagger_{x,y,z,w}(4)c_{x,y,z,w+1}(3) - c^\dagger_{x,y,z,w}(5)c_{x,y,z,w+1}(6) - c^\dagger_{x,y,z,w}(8)c_{x,y,z,w+1}(7) \\
&- c^\dagger_{x,y,z,w}(9)c_{x,y,z,w+1}(10) - c^\dagger_{x,y,z,w}(12)c_{x,y,z,w+1}(11) + c^\dagger_{x,y,z,w}(13)c_{x,y,z,w+1}(14) + c^\dagger_{x,y,z,w}(16)c_{x,y,z,w+1}(15)]\} + H.C.
\end{aligned}  \tag{S3}$$

Here, $c^\dagger_{x,y,z,w}$ and $c_{x,y,z,w}$ are the creation and annihilation operators at the ($x$, $y$, $z$, $w$)

location. Performing the Fourier transformation of $H(x, y, z, w)$, we have:

$$\begin{aligned}\tilde{H}(k_x,k_y,k_z,k_w) = &\gamma[(c_1^\dagger c_2 - c_1^\dagger c_4 + c_1^\dagger c_5 + c_1^\dagger c_9) + (c_2^\dagger c_3 + c_2^\dagger c_6 + c_2^\dagger c_{10}) + (c_3^\dagger c_4 + c_3^\dagger c_7 + c_3^\dagger c_{11}) + (c_4^\dagger c_8 + c_4^\dagger c_{12}) + \\ &(-c_5^\dagger c_6 + c_5^\dagger c_8 + c_5^\dagger c_{13}) + (-c_6^\dagger c_7 + c_6^\dagger c_{14}) + (-c_7^\dagger c_8 + c_7^\dagger c_{15}) + c_8^\dagger c_{16} + (-c_9^\dagger c_{10} + c_9^\dagger c_{12} - c_9^\dagger c_{13}) + \\ &(-c_{10}^\dagger c_{11} - c_{10}^\dagger c_{14}) + (-c_{11}^\dagger c_{12} - c_{11}^\dagger c_{15}) - c_{12}^\dagger c_{16} + (c_{13}^\dagger c_{14} - c_{13}^\dagger c_{16}) + c_{14}^\dagger c_{15} + c_{15}^\dagger c_{16}] + \\ &\lambda[e^{ik_x}(c_1^\dagger c_9 + c_2^\dagger c_{10} + c_3^\dagger c_{11} + c_4^\dagger c_{12} + c_5^\dagger c_{13} + c_6^\dagger c_{14} + c_7^\dagger c_{15} + c_8^\dagger c_{16}) + \\ &e^{ik_y}(c_1^\dagger c_5 + c_2^\dagger c_6 + c_3^\dagger c_7 + c_4^\dagger c_8 - c_9^\dagger c_{13} - c_{10}^\dagger c_{14} - c_{11}^\dagger c_{15} - c_{12}^\dagger c_{16}) + \\ &e^{ik_z}(-c_1^\dagger c_4 + c_2^\dagger c_3 + c_5^\dagger c_8 - c_6^\dagger c_7 + c_9^\dagger c_{12} - c_{10}^\dagger c_{11} - c_{13}^\dagger c_{16} + c_{14}^\dagger c_{15}) + \\ &e^{ik_w}(c_1^\dagger c_2 + c_4^\dagger c_3 - c_5^\dagger c_6 - c_8^\dagger c_7 - c_9^\dagger c_{10} - c_{12}^\dagger c_{11} + c_{13}^\dagger c_{14} + c_{16}^\dagger c_{15})] + H.C\end{aligned}$$ (S4)

The Eq. (S4) can be written in the form of: $\tilde{H}(k_x,k_y,k_z,k_w) = \vec{c}^\dagger \ddot{H}(k_{4D}) \vec{c}$, in this case, the 16-band Bloch Hamiltonian is given by:

$$H(k) = \begin{bmatrix} \cdots \end{bmatrix}$$ (S5)

The Eq. (S5) can be written in the form of Pauli matrices as presented in Eq. (1) in the main text. $H(k)$ has energies

$$E = \pm\sqrt{4(\gamma^2 + \lambda^2) + 2\gamma\lambda(\cos k_x + \cos k_y + \cos k_z + \cos k_w)},$$ (S6)

each of which is eightfold degenerate. The energy gap exists unless $\gamma/\lambda = \pm 1$. Hence, at half filling (eight electrons per cell), our proposed lattice model is an insulator. A phase transition occurs at $\gamma/\lambda = 1$ ($\gamma/\lambda = -1$), with a bulk energy gap closing at the ($\pi$, $\pi$, $\pi$, $\pi$) [(0, 0, 0, 0)] point of the BZ.

## C. Nested Wilson loop and topological invariants

In this part, we present the nested Wilson loop method for the 4[th]-order/4D hexadecapole TI. Similar to the quadrupole and octupole TIs, the topology of our proposed 4D hexadecapole TIs can also be evaluated based on the nested Wilson loop. We note that our designed hexadecapole insulator possesses eightfold occupied bands at half filling. In this case, to reveal the topology of 4D hexadecapole topological insulators, we begin the analysis by performing the first Wilson loop along the *w* direction ($W_w$), which is constructed by the Bloch states below the band-gap ($|u_k^n>$ with $n=1\ldots8$). The element of $W_w$ is $[F_{w,k}]^{m,n} = <u_{k+\Delta k_w}^m | u_k^n>$, where $\Delta k_w = (0,0,\frac{2\pi}{N_w})$ and $N_w$ is the discrete number of Wilson loop along $k_w$ direction in the first Brillouin zone (BZ). Hence, the Wilson loop is $W_w = F_{w,k+N_w\Delta k_w}\ldots F_{w,k+\Delta k_w} F_{w,k}$ and the corresponding eigenvalue problem is,

$$W_w | v_w^j> = exp[i2\pi v_w^j(k_x,k_y,k_z)] | v_w^j>. \tag{S7}$$

The Wilson loop along the $w$ direction is represented by the $8\times8$ matrix with the eigenstates being $|v_w^j>$ for $j=1\ldots8$. The first Wilson loop splits the original degenerate eightfold occupied bands into two fourfold 3D Wannier sectors, $v_w^+(k_x,k_y,k_z)$ and $v_w^-(k_x,k_y,k_z)$, separated by a Wannier gap ($\gamma/\lambda\neq\pm1$). The existence of the Wannier gap is protected by four anti-commute reflection symmetries of the Bloch Hamiltonian.

To illuminate the topology of each 3D Wannier sectors, we calculate the second nested Wilson loop within one of the 3D Wannier sectors along the z-direction. Without loss of generality, we choose $v_w^+(k_x,k_y,k_z)$. Here, we firstly discuss the method of the construction of nested Wilson loop.

The Eq. (S7) can be rewritten in the separated Wannier sector space ['+$w$'/'−$w$' corresponds to $v_w^+(k_x,k_y,k_z)/v_w^-(k_x,k_y,k_z)$ ] as:

$$W_w | v_w^{\pm w,j}> = exp[i2\pi v_w^{\pm w}(k_x,k_y,k_z)] | v_w^{\pm w,j}>. \tag{S8}$$

In this case, four Wannier states in the selected Wannier sector ('+$w$') can be constructed,

$$| W_w^{+w,j}> = \sum_{n=1}^{8} |u_k^n> [|v_w^{+w,j}>]^n \tag{S9}$$

for $j=1\ldots4$. Here, $[|v_{w,k}^{+w,j}>]^n$ is the $n$th component of $|v_w^{+w,j}>$, which is the $j$th eigenstate of the first Wilson-loop ($W_w$) in the '+$w$' Wannier sector. The second nested Wilson loop ($|\tilde{W}_z^{+w}>$, $4\times4$) along the z direction can be constructed by using the basis $|W_w^{+w,j}>$. The element of the second nested Wilson loop along the z direction is

$[F_{z,k}]^{j,i} = <W^{+w,j}_{w,k+\Delta k_z} | W^{+w,i}_{w,k}>$ (*i, j* =1, 2, 3, 4), where $\Delta k_z = (0,0,\frac{2\pi}{N_z},0)$ and $N_z$ is the discrete number of nested Wilson loop along $k_z$. Hence, the second nested Wilson loop is defined as $\tilde{W}^{+w}_z = F_{z,k+N_z\Delta k_z}...F_{z,k+\Delta k_z}F_{z,k}$. Implementing the proposed second nest Wilson loop $\tilde{W}^{+w}_z$, we have

$$\tilde{W}^{+w}_z|\eta^{+w,\pm z,j}_z> = exp[i2\pi\eta^{\pm}_z(k_x,k_y)]|\eta^{+w,\pm z,j}_z> \quad (S10)$$

with *j*=1, 2. The second nested Wilson loop further splits the fourfold 3D Wannier sector into gapped twofold 2D Wannier sector $\eta^{+w,\pm z}_z(k_x,k_y)$.

To further explore the topology of the 2D Wannier sector $\eta^{+w,\pm z}_z(k_x,k_y)$, the third nested Wilson loop within one of the 2D Wannier sectors along the y direction should be performed. Here, we choose $\eta^{+w,+z}_z(k_x,k_y)$ ('+w, +z'). Similar to Eq. (S8), the Wannier states in the selected 2D Wannier sector [$\eta^{+w,+z}_z(k_x,k_y)$] should be expressed as:

$$|W^{+w,+z,j}_z> = \sum_{n=1}^{4}|u^n_k>[|\eta^{+w,+z,j}_z>]^n \quad (S11)$$

for *j*=1, 2. $[|\eta^{+w,+z,j}_{z,k}>]^n$ is the *n*th component of the *j*th Wilson-loop eigenstate $|\eta^{+w,+z,j}_z>$. The third nested Wilson loop ($\tilde{W}^{+w,+z}_y$, 2×2) along the y direction can be constructed by using the basis $|W^{+w,+z,j}_z>$. The element of $\tilde{W}^{+w,+z}_y$ is $[F_{y,k}]^{j,i} = <\eta^{+w,+z,j}_{z,k+\Delta k_y}|\eta^{+w,+z,i}_{z,k}>$ (*i, j* =1, 2), where $\Delta k_y = (0,\frac{2\pi}{N_z},0,0)$ and $N_y$ is the discrete number of Wilson loop along $k_y$. In this case, the third nested Wilson loop is defined as $\tilde{W}^{+w,+z}_y = F_{y,k+N_y\Delta k_y}...F_{y,k+\Delta k_y}F_{y,k}$. Then, we diagonalize the third nested Wilson loop $\tilde{W}^{+w,+z}_y$,

$$\tilde{W}^{+w,+z}_y|\chi^{+w,+z,\pm y}_y> = exp[i2\pi\chi^{\pm}_y(k_x)]|\chi^{+w,+z,\pm y}_y>, \quad (S12)$$

which resolves the twofold 2D Wannier sector $\eta^{+w,+z}_z(k_x,k_y)$ into the single 1D gapped Wannier band $\chi^{\pm}_y(k_x)$. Each $\chi^{\pm}_y(k_x)$ has quantized dipole moment, indicated by a Berry phase of 0 or π, which can be calculated using the fourth nested Wilson loop [$\chi^{\pm}_y(k_x)$] along the *x* direction. The corresponding Wannier basis can be expressed as

$$|W_y^{+w,+z,+y}> = \sum_{n=1}^{2} |u_k^n>[|\chi_y^{+w,+z,+y}>]^n \tag{S13}$$

with $[|\chi_{y,k}^{+w,+z,+y}>]^n$ being the $n$th component of the third nested Wilson-loop eigenstate $|\chi_{y,k}^{+w,+z,+y}>$. The fourth nested Wilson loop ($\tilde{W}_x^{+w,+z,+y}$, $1\times1$) along $x$ can be constructed by using the basis $|\chi_y^{+w,+z,+y}>$. The element of the last nested Wilson loop along the x direction is $F_{x,k} = <\chi_{y,k+\Delta k_x}^{+w,+z,+y}|\chi_{y,k}^{+w,+z,+y}>$, where $\Delta k_x = (\frac{2\pi}{N_x},0,0,0)$ and $N_x$ is the discrete number of Wilson loop along $k_x$. This Wilson loop is associated with the Wannier-sector polarization,

$$P_x^{+w,+z,+y} = -\frac{i}{2\pi N_y N_z N_w} \sum_{kw,kz,ky} log[\tilde{W}_x^{+w,+z,+y}]. \tag{S14}$$

In our model, when the intra-cell coupling is smaller (larger) than the inter-cell coupling, the four anti-commute reflection symmetries quantizes the polarization of Wannier-sector to be 1/2 (0). It is worthy to note that the order of the nested Wilson loops is arbitrary, that is $P_x^{+w,+z,+y} = P_x^{+y,+w,+z}$. Hence, in the main text, the Wannier-sector polarization $P_x^{+w,+z,+y}$ is labelled by $P_x$.

## D. Derivation of circuit Laplacian $\tilde{J}(\omega_0,k)$ in the moment space

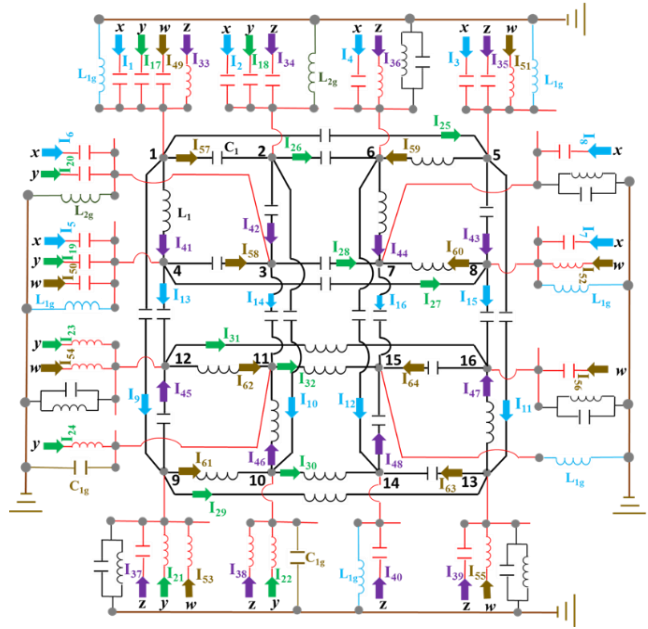

**Figure S2 | The unit cell of proposed circuits with hexadecapole moments.** The arrows mark the direction of currents.

The unit cell of our proposed circuits with hexadecapole moments is shown in Fig. S2. Here, the voltages and currents on all nodes are labeled by $V_1\ldots V_{16}$ and $I_1\ldots I_{16}$, respectively. $I_{G1}$ to $I_{G16}$ represent the currents flowing out from the nodes 1 to 16. $k_x$, $k_y$, $k_z$, and $k_w$ denote the phase of Block wave vector propagating in the $x$, $y$, z and $w$ directions, respectively. We apply the Kirchhoff's current Law to the sixteen nodes, and have:

$$I_1 + I_{17} + I_{33} + I_{49} = I_9 + I_{25} + I_{41} + I_{57} + I_{G_1},$$
$$I_2 + I_{18} + I_{34} + I_{57} = I_{10} + I_{26} + I_{42} + I_{49}e^{-ik_w} + I_{G_2},$$
$$I_6 + I_{20} + I_{42} + I_{58} = I_{14} + I_{28} + I_{34}e^{-ik_z} + I_{50}e^{-ik_w} + I_{G_3},$$
$$I_5 + I_{19} + I_{41} + I_{50} = I_{13} + I_{27} + I_{33}e^{-ik_z} + I_{58} + I_{G_4},$$
$$I_3 + I_{25} + I_{35} + I_{51} = I_{11} + I_{17}e^{-ik_y} + I_{43} + I_{59} + I_{G_5},$$
$$I_4 + I_{26} + I_{36} + I_{59} = I_{12} + I_{18}e^{-ik_y} + I_{44} + I_{51}e^{-ik_w} + I_{G_6},$$
$$I_8 + I_{28} + I_{44} + I_{60} = I_{16} + I_{20}e^{-ik_y} + I_{36}e^{-ik_z} + I_{52}e^{-ik_w} + I_{G_7},$$
$$I_7 + I_{27} + I_{43} + I_{52} = I_{15} + I_{19}e^{-ik_y} + I_{35}e^{-ik_z} + I_{60} + I_{G_8},$$
$$I_9 + I_{21} + I_{37} + I_{53} = I_1 e^{-ik_x} + I_{29} + I_{45} + I_{61} + I_{G_9},$$
$$I_{10} + I_{22} + I_{38} + I_{61} = I_2 e^{-ik_x} + I_{30} + I_{46} + I_{53}e^{-ik_z} + I_{G_{10}},$$
$$I_{14} + I_{24} + I_{46} + I_{62} = I_6 e^{-ik_x} + I_{32} + I_{38}e^{-ik_z} + I_{54}e^{-ik_w} + I_{G_{11}},$$
$$I_{13} + I_{23} + I_{45} + I_{54} = I_5 e^{-ik_x} + I_{31} + I_{37}e^{-ik_z} + I_{62} + I_{G_{12}},$$
$$I_{11} + I_{29} + I_{39} + I_{55} = I_3 e^{-ik_x} + I_{21}e^{-ik_y} + I_{47} + I_{63} + I_{G_{13}},$$
$$I_{12} + I_{30} + I_{40} + I_{63} = I_4 e^{-ik_x} + I_{22}e^{-ik_y} + I_{48} + I_{55}e^{-ik_w} + I_{G_{14}}, \qquad (S15)$$
$$I_{16} + I_{32} + I_{48} + I_{64} = I_8 e^{-ik_x} + I_{24}e^{-ik_y} + I_{40}e^{-ik_z} + I_{56}e^{-ik_w} + I_{G_{15}},$$
$$I_{15} + I_{31} + I_{47} + I_{56} = I_7 e^{-ik_x} + I_{23}e^{-ik_y} + I_{39}e^{-ik_z} + I_{64} + I_{G_{16}}.$$

According to the current direction given in Fig. S2, we can write the current that flows out of each node a

$$I_1 = -i\omega C_2\left(V_9 e^{ik_x} - V_1\right) - i\omega C_2\left(V_5 e^{ik_y} - V_1\right) - \frac{\left(V_4 e^{ik_z} - V_1\right)}{i\omega L_2} - i\omega C_2\left(V_2 e^{ik_w} - V_1\right)$$
$$+ i\omega C_1(V_1 - V_9) + i\omega C_1(V_1 - V_5) + \frac{(V_1 - V_4)}{i\omega L_1} + i\omega C_1(V_1 - V_2) + \frac{2V_1}{i\omega L_1} + \frac{2V_1}{i\omega L_2},$$

$$I_2 = -i\omega C_2\left(V_{10} e^{ik_x} - V_2\right) - i\omega C_2\left(V_6 e^{ik_y} - V_2\right) - i\omega C_2\left(V_3 e^{ik_z} - V_2\right) - i\omega C_1(V_1 - V_2)$$
$$+ i\omega C_1(V_2 - V_{10}) + i\omega C_1(V_2 - V_6) + i\omega C_1(V_2 - V_3) + i\omega C_2\left(V_2 - V_1 e^{-ik_w}\right) + \frac{4V_2}{i\omega L_1} + \frac{4V_2}{i\omega L_2},$$

$$I_3 = -i\omega C_2\left(V_{11} e^{ik_x} - V_3\right) - i\omega C_2\left(V_7 e^{ik_y} - V_3\right) - i\omega C_1(V_2 - V_3) - i\omega C_1(V_4 - V_3)$$
$$+ i\omega C_1(V_3 - V_{11}) + i\omega C_1(V_3 - V_7) + i\omega C_2\left(V_3 - V_2 e^{-ik_z}\right) + i\omega C_2\left(V_3 - V_4 e^{-ik_w}\right) + \frac{4V_3}{i\omega L_1} + \frac{4V_3}{i\omega L_2},$$

$$I_4 = -i\omega C_2\left(V_{12}e^{ik_x} - V_4\right) - i\omega C_2\left(V_8 e^{ik_y} - V_4\right) - \frac{(V_1 - V_4)}{i\omega L_1} - i\omega C_2\left(V_3 e^{ik_w} - V_4\right)$$
$$+ i\omega C_1(V_4 - V_{12}) + i\omega C_1(V_4 - V_8) + \frac{V_4 - V_1 e^{-ik_z}}{i\omega L_2} + i\omega C_1(V_4 - V_3) + \frac{2V_4}{i\omega L_1} + \frac{2V_4}{i\omega L_2},$$

$$I_5 = -i\omega C_2\left(V_{13}e^{ik_x} - V_5\right) - i\omega C_1(V_1 - V_5) - i\omega C_2\left(V_8 e^{ik_z} - V_5\right) - \frac{V_6 e^{ik_w} - V_5}{i\omega L_2}$$
$$+ i\omega C_1(V_5 - V_{13}) + i\omega C_2\left(V_5 - V_1 e^{-ik_y}\right) + i\omega C_1(V_5 - V_8) + \frac{V_5 - V_6}{i\omega L_1} + \frac{2V_5}{i\omega L_1} + \frac{2V_5}{i\omega L_2},$$

$$I_6 = -i\omega C_2\left(V_{14}e^{ik_x} - V_6\right) - i\omega C_1(V_2 - V_6) - \frac{V_7 e^{ik_z} - V_6}{i\omega L_2} - \frac{V_5 - V_6}{i\omega L_1}$$
$$+ i\omega C_1(V_6 - V_{14}) + i\omega C_2\left(V_6 - V_2 e^{-ik_y}\right) + \frac{V_6 - V_7}{i\omega L_1} + \frac{V_6 - V_5 e^{-ik_w}}{i\omega L_2} + \frac{V_6}{i\omega L_1} + i\omega C_1 V_6,$$

$$I_7 = -i\omega C_2\left(V_{15}e^{ik_x} - V_7\right) - i\omega C_1(V_3 - V_7) - \frac{V_6 - V_7}{i\omega L_1} - \frac{V_8 - V_7}{i\omega L_1}$$
$$+ i\omega C_1(V_7 - V_{15}) + i\omega C_2\left(V_7 - V_3 e^{-ik_y}\right) + \frac{V_7 - V_6 e^{-ik_z}}{i\omega L_2} + \frac{V_7 - V_8 e^{-ik_w}}{i\omega L_2} + \frac{V_7}{i\omega L_1} + i\omega C_1 V_7,$$

$$I_8 = -i\omega C_2\left(V_{16}e^{ik_x} - V_8\right) - i\omega C_1(V_4 - V_8) - i\omega C_1(V_5 - V_8) - \frac{V_7 e^{ik_w} - V_8}{i\omega L_2}$$
$$+ i\omega C_1(V_8 - V_{16}) + i\omega C_2\left(V_8 - V_4 e^{-ik_y}\right) + i\omega C_2(V_8 - V_5 e^{-ik_z}) + \frac{V_8 - V_7}{i\omega L_1} + \frac{2V_8}{i\omega L_1} + \frac{2V_8}{i\omega L_2},$$

$$I_9 = -i\omega C_1(V_1 - V_9) - \frac{V_{13}e^{ik_y} - V_9}{i\omega L_2} - i\omega C_2(V_{12}e^{ik_z} - V_9) - \frac{V_{10}e^{ik_w} - V_9}{i\omega L_2}$$
$$+ i\omega C_2\left(V_9 - V_1 e^{-ik_x}\right) + \frac{V_9 - V_{13}}{i\omega L_1} + i\omega C_1(V_9 - V_{12}) + \frac{V_9 - V_{10}}{i\omega L_1} + \frac{V_9}{i\omega L_1} + i\omega C_1 V_9,$$

$$I_{10} = -i\omega C_1(V_2 - V_{10}) - \frac{V_{14}e^{ik_y} - V_{10}}{i\omega L_2} - \frac{V_{11}e^{ik_z} - V_{10}}{i\omega L_2} - \frac{V_9 - V_{10}}{i\omega L_1}$$
$$+ i\omega C_2\left(V_{10} - V_2 e^{-ik_x}\right) + \frac{V_{10} - V_{14}}{i\omega L_1} + \frac{V_{10} - V_{11}}{i\omega L_1} + \frac{V_{10} - V_9 e^{-ik_w}}{i\omega L_2} + 2i\omega C_1 V_{10} + 2i\omega C_2 V_{10},$$

$$I_{11} = -i\omega C_1(V_3 - V_{11}) - \frac{V_{15}e^{ik_y} - V_{11}}{i\omega L_2} - \frac{V_{10} - V_{11}}{i\omega L_1} - \frac{V_{12} - V_{11}}{i\omega L_1}$$
$$+ i\omega C_2\left(V_{11} - V_3 e^{-ik_x}\right) + \frac{V_{11} - V_{15}}{i\omega L_1} + \frac{V_{11} - V_{10}e^{-ik_z}}{i\omega L_2} + \frac{V_{11} - V_{12}e^{-ik_w}}{i\omega L_2} + 2i\omega C_1 V_{11} + 2i\omega C_2 V_{11},$$

$$I_{12} = -i\omega C_1(V_4 - V_{12}) - \frac{V_{16}e^{ik_y} - V_{12}}{i\omega L_2} - i\omega C_1(V_9 - V_{12}) - \frac{V_{11}e^{ik_w} - V_{12}}{i\omega L_2}$$
$$+ i\omega C_2\left(V_{12} - V_4 e^{-ik_x}\right) + \frac{V_{12} - V_{16}}{i\omega L_1} + i\omega C_2(V_{12} - V_9 e^{-ik_z}) + \frac{V_{12} - V_{11}}{i\omega L_1} + \frac{V_{12}}{i\omega L_1} + i\omega C_1 V_{12},$$

$$I_{13} = -i\omega C_1(V_5 - V_{13}) - \frac{V_9 - V_{13}}{i\omega L_1} - \frac{V_{16}e^{ik_z} - V_{13}}{i\omega L_2} - i\omega C_2(V_{14}e^{ik_w} - V_{13})$$
$$+ i\omega C_2\left(V_{13} - V_5 e^{-ik_x}\right) + \frac{V_{13} - V_9 e^{-ik_y}}{i\omega L_2} + \frac{V_{13} - V_{16}}{i\omega L_1} + i\omega C_1(V_{13} - V_{14}) + \frac{V_{13}}{i\omega L_1} + i\omega C_1 V_{13},$$

$$I_{14} = -i\omega C_1(V_6 - V_{14}) - \frac{V_{10} - V_{14}}{i\omega L_1} - i\omega C_2(V_{15}e^{ik_z} - V_{14}) - i\omega C_1(V_{13} - V_{14})$$
$$+ i\omega C_2\left(V_{14} - V_6 e^{-ik_x}\right) + \frac{V_{14} - V_{10}e^{-ik_y}}{i\omega L_2} + i\omega C_1(V_{14} - V_{15}) + i\omega C_2(V_{14} - V_{13}e^{-ik_w}) + \frac{2V_{14}}{i\omega L_1} + \frac{2V_{14}}{i\omega L_2},$$

$$I_{15} = -i\omega C_1(V_7 - V_{15}) - \frac{V_{11} - V_{15}}{i\omega L_1} - i\omega C_1(V_{14} - V_{15}) - i\omega C_1(V_{16} - V_{15}) \quad\quad (S16)$$
$$+ i\omega C_2\left(V_{15} - V_7 e^{-ik_x}\right) + \frac{V_{15} - V_{11}e^{-ik_y}}{i\omega L_2} + i\omega C_2(V_{15} - V_{14}e^{-ik_z}) + i\omega C_2(V_{15} - V_{16}e^{-ik_w}) + \frac{2V_{15}}{i\omega L_1} + \frac{2V_{15}}{i\omega L_2},$$

$$I_{16} = -i\omega C_1(V_8 - V_{16}) - \frac{V_{12} - V_{16}}{i\omega L_1} - \frac{V_{13} - V_{16}}{i\omega L_1} - i\omega C_2(e^{ik_w}V_{15} - V_{16})$$
$$+ i\omega C_2\left(V_{16} - V_8 e^{-ik_x}\right) + \frac{V_{16} - V_{12}e^{-ik_y}}{i\omega L_2} + \frac{V_{16} - V_{13}e^{-ik_z}}{i\omega L_2} + i\omega C_1(V_{16} - V_{15}) + \frac{V_{16}}{i\omega L_1} + i\omega C_1 V_{16}.$$

Writing Eqs. (S16) in the form of $\vec{I} = \tilde{J}(\omega, k_{4D})\vec{V}$, the circuit Laplacian in the moment space can be expressed as:

$$\tilde{J}(\omega, k) = i\omega \begin{vmatrix} \cdots \end{vmatrix}$$

(S17)

When the driving frequency equals to $\omega=\omega_0=1/(L_1C_1)^{1/2}=1/(L_2C_2)^{1/2}$, the circuit Laplacian in the moment space becomes:

$$\tilde{J}(\omega_0, k_{4D}) = -i\sqrt{C_1/L_1}[(1+\kappa\cos k_x)\Gamma_1 + \kappa\sin k_x\Gamma_2 + (1+\kappa\cos k_y)\Gamma_3 + \kappa\sin k_y\Gamma_4 \\ + (1+\kappa\cos k_z)\Gamma_5 + \kappa\sin k_z\Gamma_6 + (1+\kappa\cos k_w)\Gamma_7 + \kappa\sin k_w\Gamma_8]$$

(S18)

Here, $\Gamma_i$ (i=1, 2, 3, 4, 5, 6, 7, 8) possesses the same form used in Eq. (1), $C_2=\kappa C_1$ and $L_1=\kappa L_2$. The circuit Laplacian $\tilde{J}(\omega_0, k_{4D})$, relating the input currents in terms of the potentials, replaces the role of the Hamiltonian in determining the spectrum relevant to the impedance. In this case, our designed circuits Laplacian Eq. (S18) possesses four anti-commute reflection symmetries, resulting in topologically protected in-gap corner state.

### E. Grounding of the finite electric circuit

To fulfill the lattice Hamiltonian of hexadecapole TIs by a topolectrical circuit Laplacian, the diagonal elements of open circuit Laplacian at the resonant frequency $\omega_0$ should be eliminated by a suitable choice of the grounding. This can be easily realized by making the fact that inductivities and capacitances enter the circuit Laplacian with opposite sign, that is,

$$J_{ab}(\omega) = i\omega C_{ab} - \frac{i}{\omega}\frac{1}{L_{ab}}. \tag{S19}$$

Hence, the contribution of inductivities (capacitances) at the fix node can be cancelled by grounding matched capacitances (inductivities). In this case, the groundings on the circuit sites located at 4D/3D bulks, 2D surfaces, 1D hinges and 0D corners are different and should be suitably designed resulting from the fact that different amount of linked capacitors/inductors on these nodes. We present the grounding patterns in Tables 1-9, which correspond to the actual sample composed of nine PCBs.

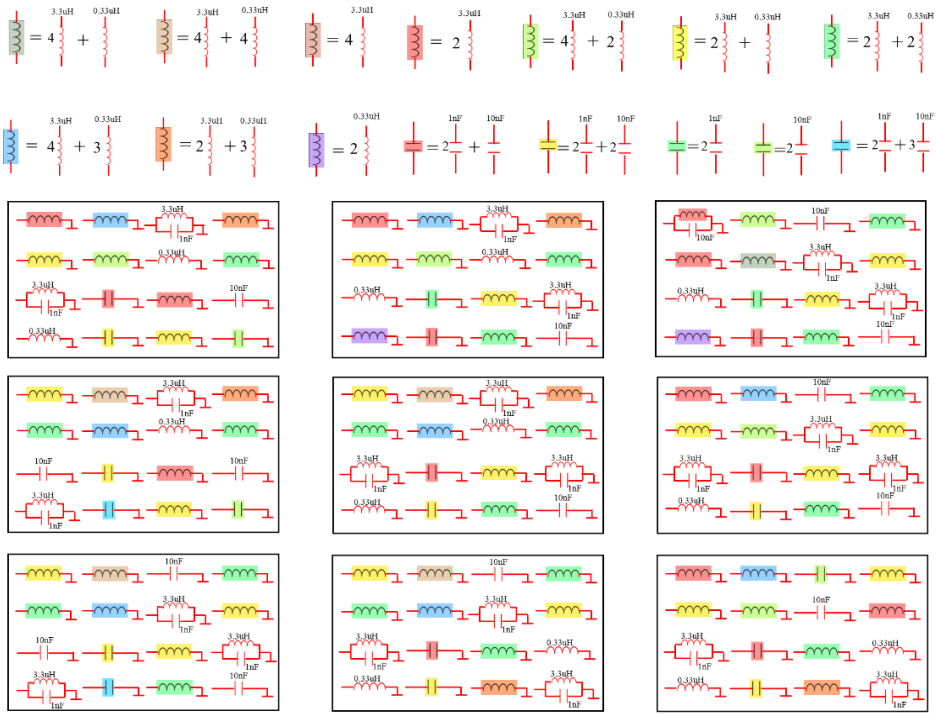

Table 1

**Table 2**

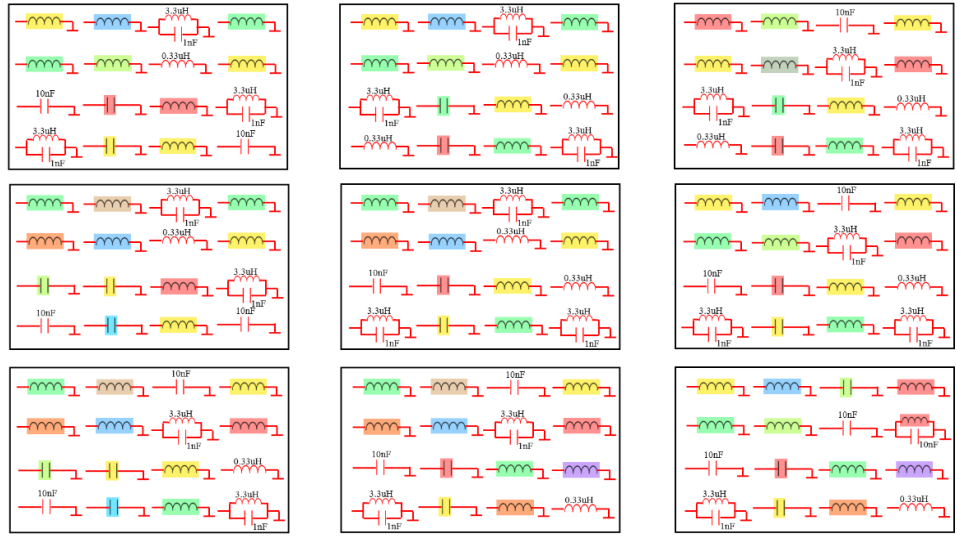

**Table 3**

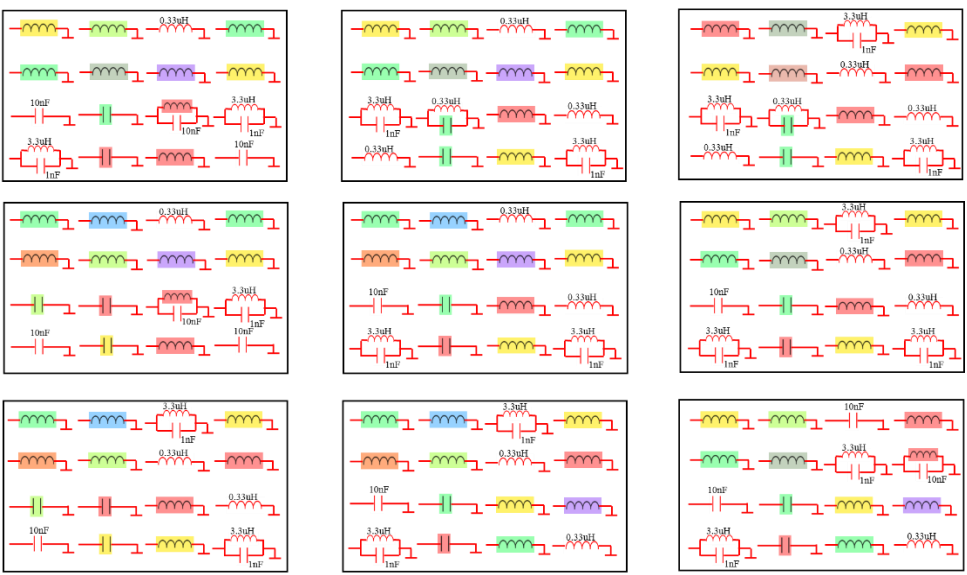

**Table 4**

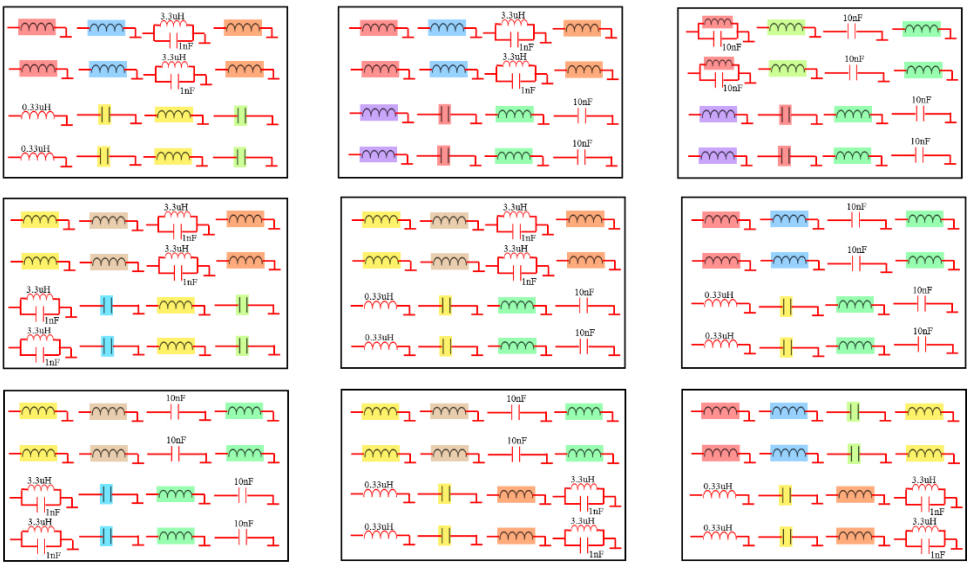

**Table 5**

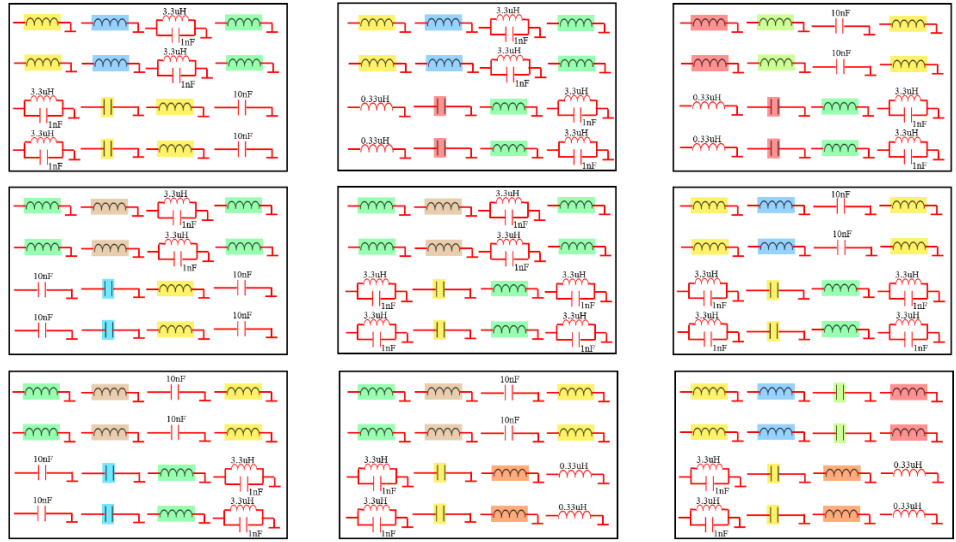

**Table 6**

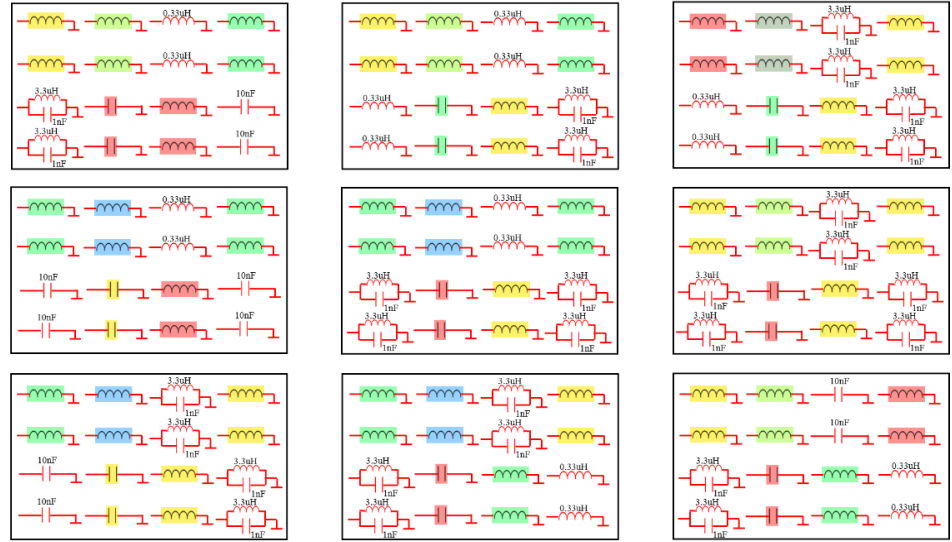

**Table 7**

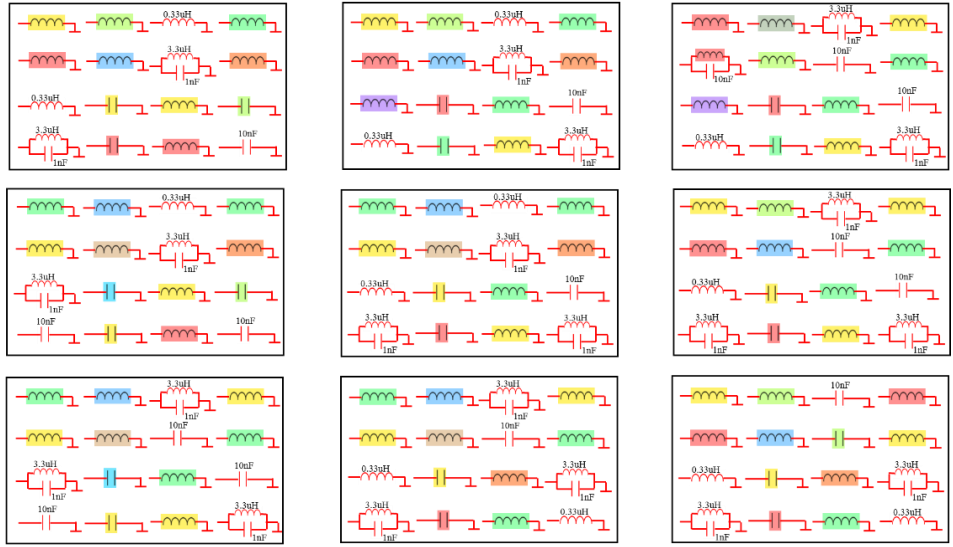

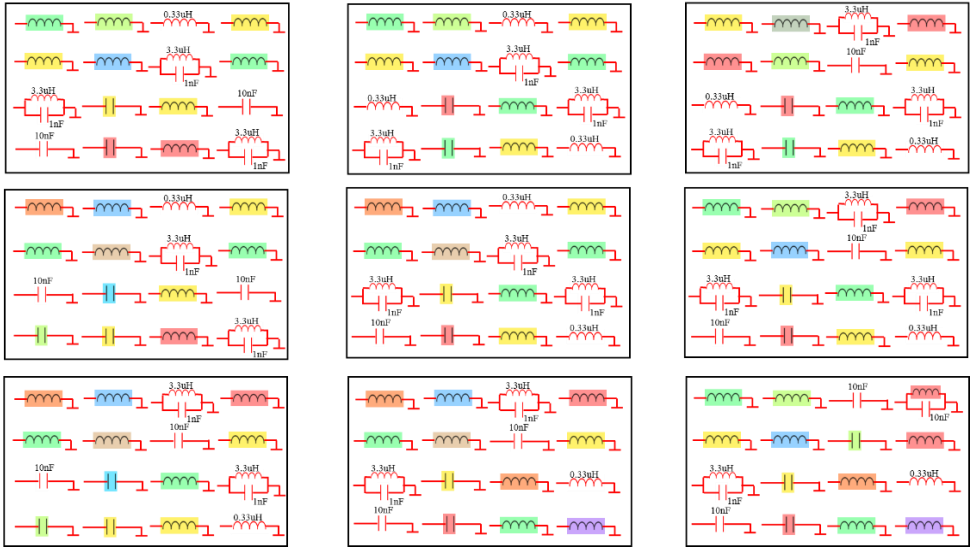

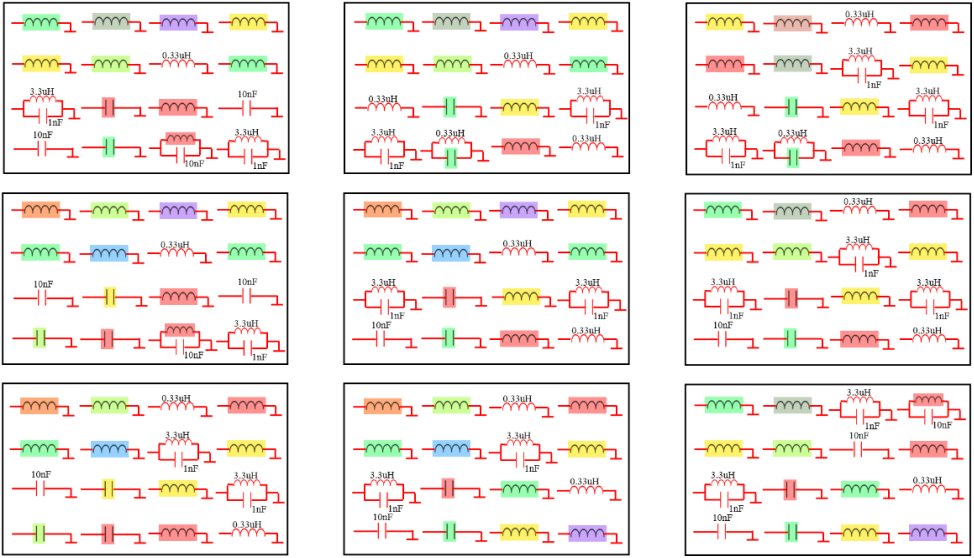

## F. Numerical analysis of the robustness for corner states

The most important property of the topologically protected corner state is that it is robust against disorders. In this part, we present numerical results of tight-binding calculations as well as actual circuit simulations on the robustness of the corner states. Here, two types of perturbation are considered. (1) The first one is related to the resonance frequency shifts on the four sites linked to the corner location (perturbation 1). (2) The second one is resonance frequency shift on the corner site (perturbation 2). The frequency shift can be easily realized by changing the grounding on the selected sites.

As shown in Fig. S3(a), we plots the eigen-spectrum of the tight-binding circuit Laplacian (3×3×3×3) with perturbation 1. It is clearly shown that the corner states is almost unchanged, demonstrating the robustness of corner states. For perturbation 2, the significant frequency shift of corner state (3×3×3×3 open circuit) appears, as shown in Fig. S3b. Except for the tight-binding lattice model calculations, we also perform circuit simulations by adding different values of capacitor on the grounding part of the perturbed sites. The simulation results under these two types of perturbations are shown in Figs. S3(c) and 3(d). The black, pink, green and blue lines correspond to the case with the capacitance of the grounded capacitor being 1nF, 3.3nF, 6.6nF and 10nF, respectively. We note that the corner state is almost unaffected by perturbing the four sites linked to the corner. However, the significant frequency shift appears when the corner site is perturbed.

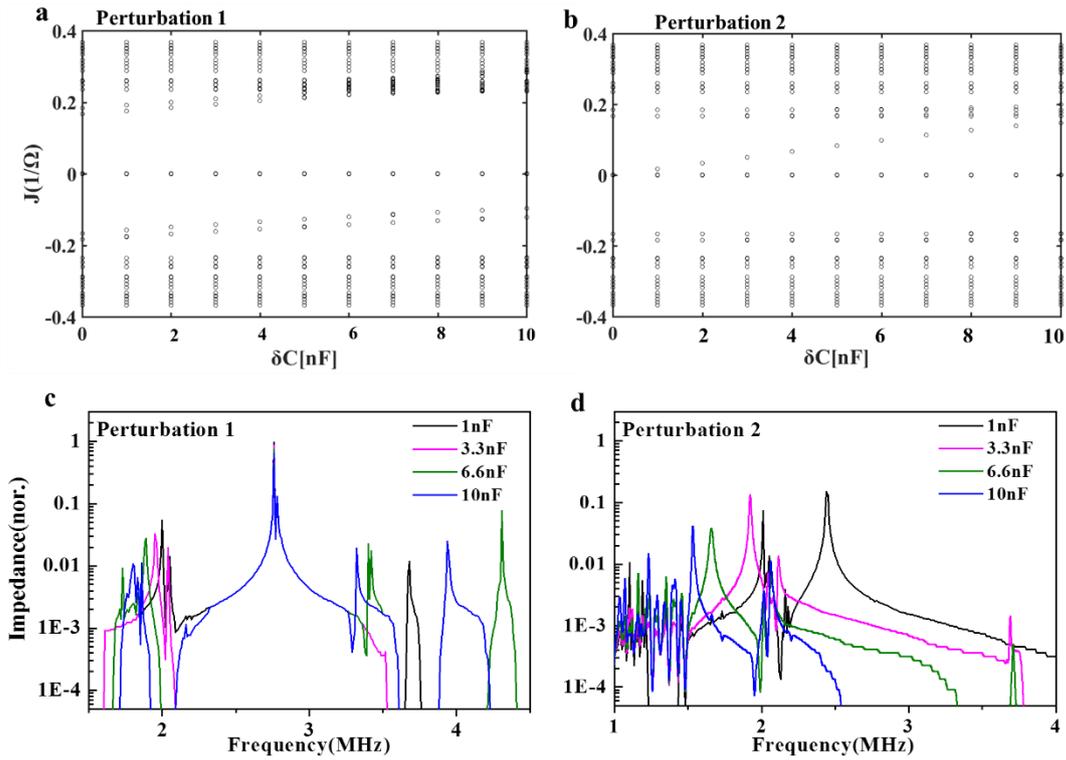

**Figure S3 | Robustness of corner states. a and b,** Eigen-spectrum of the tight-binding lattice model (3×3×3×3) with perturbation 1 and 2, respectively. **c and d,** The circuit simulation results under two types of perturbations.